\pdfoutput=1

\documentclass[sigconf]{acmart}

\usepackage{enumitem}
\usepackage{tikz-qtree}
\usepackage{subcaption}

\newcommand{\sect}{\textsection}

\AtBeginDocument{%
  \providecommand\BibTeX{{%
    \normalfont B\kern-0.5em{\scshape i\kern-0.25em b}\kern-0.8em\TeX}}}

\acmConference[AIChallengeIoT 2019]{First International Workshop on Challenges in Artificial Intelligence and Machine Learning for Internet of Things}{November 10-13, 2019}{New York, NY}
\setcopyright{acmcopyright}

\begin{document}
\title{Challenges of Privacy-Preserving Machine Learning in IoT}

\author{Mengyao Zheng}
\authornote{The first two authors contributed equally to this research. This work was completed when they were with Xi'an Jiaotong-Liverpool University and visiting Nanyang Technological University.}
\affiliation{
	\institution{Nanyang Technological University}
}

\author{Dixing Xu}
\authornotemark[1]
\affiliation{
	\institution{Nanyang Technological University}
}

\author{Linshan Jiang}
\affiliation{
  \institution{Nanyang Technological University}
}

\author{Chaojie Gu}
\affiliation{
	\institution{Nanyang Technological University}
}

\author{Rui Tan}
\affiliation{
	\institution{Nanyang Technological University}
}

\author{Peng Cheng}
\affiliation{
        \institution{Zhejiang University}
}

\renewcommand{\shortauthors}{Zheng and Xu, et al.}

\begin{abstract}
The Internet of Things (IoT) will be a main data generation infrastructure for achieving better system intelligence. However, the extensive data collection and processing in IoT also engender various privacy concerns. This paper provides a taxonomy of the existing privacy-preserving machine learning approaches developed in the context of cloud computing and discusses the challenges of applying them in the context of IoT. Moreover, we present a privacy-preserving inference approach that runs a lightweight neural network at IoT objects to obfuscate the data before transmission and a deep neural network in the cloud to classify the obfuscated data. Evaluation based on the MNIST dataset shows satisfactory performance.
\end{abstract}




\copyrightyear{2019}
\acmYear{2019}
\acmConference[AIChallengeIoT'19]{First International Workshop on Challenges in Artificial
	Intelligence and Machine Learning }{November 10--13, 2019}{New York, NY, USA}
\acmBooktitle{First International Workshop on Challenges in Artificial Intelligence and Machine
	Learning (AIChallengeIoT'19), November 10--13, 2019, New York, NY, USA}
\acmPrice{15.00}
\acmDOI{10.1145/3363347.3363357}
\acmISBN{978-1-4503-7013-4/19/11}

\maketitle

\section{Introduction}

With the advances of sensing and communication technologies, the Internet of Things (IoT) will become a main data generation infrastructure in the future. The drastically increasing amount of data generated by IoT will create unprecedented opportunities for various novel applications powered by machine learning (ML). However, various system challenges need to be addressed to implement the envisaged intelligent IoT.

IoT in nature is a distributed system consisting of heterogeneous nodes with distinct sensing, computation, and communication capabilities. Specifically, it consists of massive mote-class sensors deeply embedded in the physical world, personal devices that move with people, widespread network edge devices such as wireless access points, as well as the cloud backend. 
Implementing the fabric of IoT and ML faces the following two key challenges:\vspace{-0.2em}
\begin{itemize}[itemsep= 5 pt,topsep = 5 pt, leftmargin =8pt]
\item {\bf Separation of data sources and ML computation power:} Most IoT data will be generated by the end devices that often have limited computation resources, while the computation power needed by ML model training and execution will be located at the edge devices and in the cloud. Besides, the communication channels between the end devices and the edge/cloud are often constrained, in that they are limited in bandwidth, intermittent, and of long delays.
\item {\bf Privacy preservation:} As the end devices can be deeply embedded in people's private space and time, the data generated by them will contain privacy-sensitive information. To gain wide acceptance, the IoT-ML fabric must respect the human users' privacy. The lack of privacy preservation may even go against the recent legislation such as the General Data Protection Regulation in European Union. However, privacy preservation often presents substantial challenges to the system design.
\end{itemize}

Privacy-preserving ML has received extensive research in the context of cloud computing. Thus, it is of great interest to investigate whether the existing solutions can be applied in the context of IoT. To this end, this paper provides a taxonomy of the existing privacy-preserving ML approaches, which are classified into two categories: {\em privacy-preserving training} and {\em privacy-preserving inference}. The former, which has received considerable research attention, is comprised of {\em parameter transmission-based} and {\em data transmission-based} approaches. Each of them can be further divided into multiple sub-categories. In contrast, our literature survey shows that less research work concentrates on privacy-preserving inference. Since the computation and communication overheads are the key considerations in the design of IoT systems, we tentatively label the existing privacy-preserving ML approaches with {\em high-overhead}, {\em medium-overhead}, and {\em low-overhead} regarding computation complexity and {\em iterative communication}, {\em data swelling}, and {\em data retention/compression} regarding communication overhead. ML for IoT should be of low-overhead and data retention/compression.

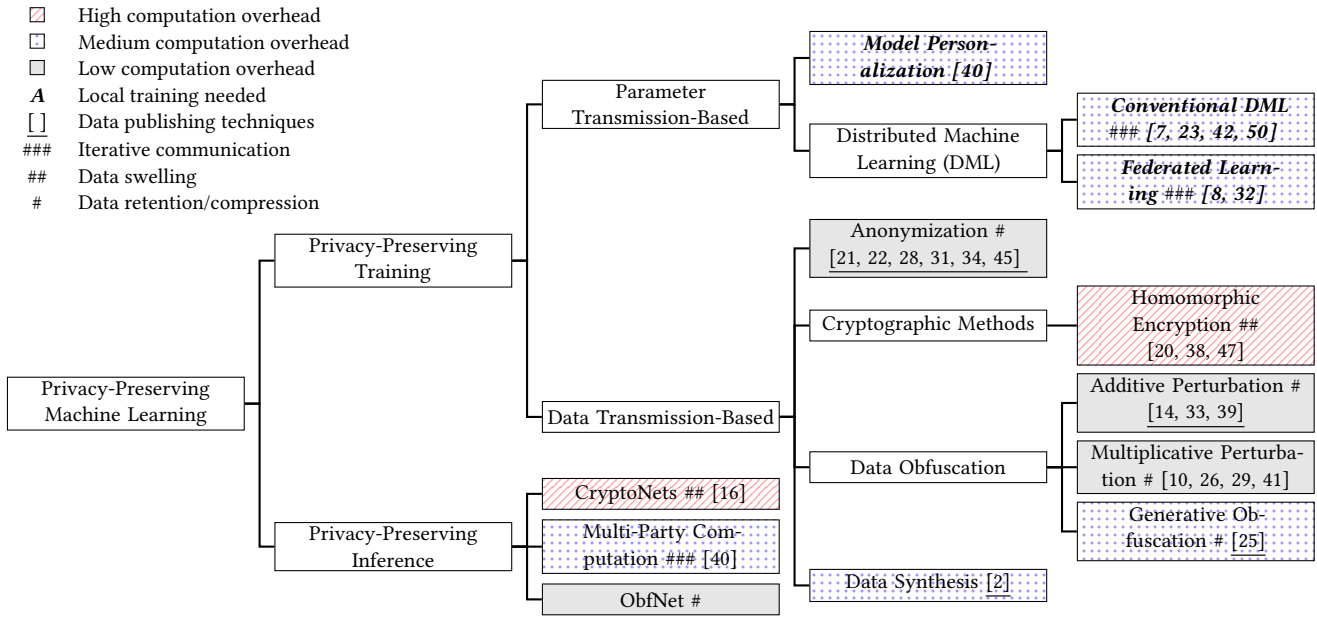
\begin{figure*} 

\usetikzlibrary{trees}
\usetikzlibrary{patterns}
	
	\begin{tikzpicture}[grow'=right,level distance=1.4in]

	\tikzset{edge from parent/.style= 
		{thick, draw, edge from parent fork right},
		every tree node/.style={draw, font=\small,minimum size=4.1mm, inner sep=0.5mm, draw, minimum width=1in,text width=1.2in,align=center}}
	
	\tikzstyle{big}  = [draw, rectangle,pattern=north east lines, pattern color=red!40]

	\tikzstyle{med} =[draw, rectangle, pattern=dots,pattern color=blue!40]
	\tikzstyle{small}   = [draw, fill=black!10]

	\tikzset{level 1/.style={sibling distance=-0.65in}}
	\tikzset{level 2/.style={sibling distance=0.05in}}
	\tikzset{level 3+/.style={sibling distance=0.04in}}
	
	\Tree 
	[.{Privacy-Preserving Machine Learning}
	[.{Privacy-Preserving Training}
	[.{Parameter Transmission-Based} 
	[.\node[med]{\textit{\textbf{Model Personalization \cite{servia2018privacy} } } }; ]
	[.{Distributed Machine Learning (DML)  }
	[.\node[med]{\textit{\textbf{ Conventional DML $\#\#\#$  \cite{hamm2015crowd,shokri2015privacy,boyd2011distributed,zinkevich2010parallelized} } } }; ]
	[.\node[med]{\textit{\textbf{ Federated Learning $\#\#\# $  \cite{mcmahan2016communication,brendan2017learning}}}};	]
	]  
	]
	[.{Data Transmission-Based } 
	[.\node[small]{ Anonymization \# {\underline{\cite{sweeney2002k,machanavajjhala2006diversity,li2007t,gross2006model,newton2005preserving,gross2005integrating} 
			} 
		}
	};
	]
	[.{Cryptographic Methods}
	[.\node[big]{ Homomorphic Encryption \#\# \\ \cite{graepel2012ml,zhan2005privacy,qi2008efficient}};
	]
	]
	[.{Data Obfuscation}
	[.\node[small]{ Additive Perturbation \# \\ \underline {\cite{dwork2006calibrating,mcsherry2007mechanism,roth2010interactive}}};
	]
	[.\node[small]{ Multiplicative Perturbation \# \cite{liu2012cloud,jiang2019lightweight,shen2018privacy,chen2005privacy}};
	]
	[.\node[med]{Generative Obfuscation \#  \underline{\cite{huang2017context}}} ;
	]
	]
	[.\node[med] {Data Synthesis \underline{\cite{acs2018differentially}} };
	]	
	]
	]
	[.{Privacy-Preserving Inference}
	[.\node[big]{ CryptoNets \#\# \cite{gilad2016cryptonets} };  ]
	[.\node[med]{ Multi-Party Computation $\#\#\#$ \cite{servia2018privacy} }; ]
	[.\node[small]{ObfNet \#};
	]
	] 
	]
	
	\node[] at (0.8,3.9) 
	{ \small
		\begin{tabular}{cl}
	
		\tikz\node[big] {}; & High computation overhead \\
		\tikz\node[med] {}; & Medium computation overhead \\
		\tikz\node[small] {}; & Low computation overhead\\
		\textit{\textbf{A}} & Local training needed\\
		\underline{[ ]} & Data publishing techniques\\
		\#\#\#& Iterative communication\\
		\#\# & Data swelling\\
		\# & Data retention/compression
		\end{tabular}
	};

	\end{tikzpicture}
	
	\caption{The hierachical taxonomy of privacy-preserving machine learning approaches.}
	\label{Fig.main2}
	 
\end{figure*}

From our survey, a number of privacy-preserving training approaches with various privacy protection objectives \cite{sweeney2002k,machanavajjhala2006diversity,li2007t,gross2006model,newton2005preserving,gross2005integrating,liu2012cloud,jiang2019lightweight,shen2018privacy,chen2005privacy,dwork2006calibrating,mcsherry2007mechanism,roth2010interactive} can be labeled low-overhead and data retention/compression. In contrast, the existing privacy-preserving inference approaches have high computation or communication overheads. {Thus, we think that privacy-preserving inference for IoT should receive more research attention, since many IoT applications leverage pre-trained ML models for inference instead of training models from scratch.} Thus, in the second part of this paper, we present a lightweight and voluntary privacy-preserving inference approach that is suitable for resource-constrained IoT objects. The approach is as follows. Given a well trained deep model that will be executed by the edge/cloud for inference, the approach trains a few dense layers such that the concatenation of these layers and the given deep model still yields satisfactory inference accuracy. At run time, an IoT end device executes these dense layers to obfuscate the inference data sample and sends the result to the edge/cloud for inference. Our approach is voluntary in that the deep model at the edge/cloud admits both original and obfuscated data samples. {Each IoT end device in the system can choose to execute the dense layers for obfuscation or simply send the original data for inference. This design accommodates the end devices that cannot perform the obfuscation due to say limited computing capability.} The evaluation based on the MNIST dataset \cite{lecun1998mnist} of handwritten digits shows that our {obfuscation} approach well protects the confidentiality of the raw form of the data samples and maintains the inference accuracy.

The remainder of this paper is organized as follows. \sect\ref{sec2} presents the taxonomy of the existing privacy-preserving ML approaches and their limitations in the context of IoT. \sect\ref{sec:ourapproach} presents our lightweight and voluntary privacy-preserving inference approach and evaluation result. \sect\ref{sec:con} concludes this paper and discusses future work.

\section{Existing Approaches \& Limitations}
\label{sec2}

Fig. \ref{Fig.main2} illustrates the taxonomy of the existing privacy-preserving ML schemes. The nodes in a privacy-preserving ML system often have two roles: {\em participant} and {\em coordinator}. The participants are often the data generators (e.g., smartphones), whereas the coordinator (e.g., a cloud server) orchestrates the ML process. Since ML has two phases, i.e., training and inference, we classify the existing approaches into two groups at the top level. The {\em privacy-preserving training} schemes (\sect\ref{PPLS}) aim to learn a global ML model or multiple local ML models from disjoint local datasets which, if aggregated, would provide more useful/precise knowledge. Thus, the primary objective of privacy protection is to preserve the privacy of the data used for building an ML model in the training phase. Differently, {\em privacy-preserving inference} schemes (\sect\ref{PP schemes for Inference}) focus on the scenario where a global ML model at the coordinator has been trained and the participants transmit the unlabeled data to the coordinator for inference. The aim is to protect the privacy of the input data in the inference phase and maintain the inference accuracy. 

{Privacy-preserving training} schemes can be further classified based on whether the privacy-sensitive training data samples are transmitted or only the model parameters are transmitted for model training. Parameter transmission-based approaches (\sect\ref{DML}) include distributed machine learning and model personalization. The approaches that need to transmit local data samples (\sect\ref{Data Transmission-based Approaches}) can be classified into {\em anonymization}, {\em cryptographic methods}, {\em obfuscation}, and {\em data synthesis} according to the processing made on the training data. %
(1) Anonymization approaches de-identify data records but do not change the data of interest for model training. %
(2) Cryptographic methods apply cryptographic primitives to encrypt the data transmitted. %
(3) Obfuscation methods transform the training data vectors through additive perturbation, multiplicative perturbation, and generative obfuscation. %
(4) Data synthesis generates a new dataset that resembles the original dataset. %
Note that some of the data transmission-based approaches for {privacy-preserving training} are {\em data publishing} techniques, which focus on the proper sharing of the data or query results and in general do not explicitly address the problem of ML model training. {Our taxonomy includes them for the completeness of the related work review.}

Most existing privacy-preserving ML approaches were designed in the context of cloud computing. % 
The {participants} are often resource-rich nodes from smartphones to cloud servers. In particular, the overhead of communications is not a key concern due to the availability of high-speed connections (e.g., wireline networks and 4G cellular networks). %
Differently, in the context of IoT, the {participants} are often resource-constrained devices. Moreover, the communication links among them are generally constrained. Therefore, in the review of the {privacy-preserving training} (\sect\ref{PPLS}) and {inference} (\sect\ref{PP schemes for Inference}) approaches, we will qualitatively discuss their computation overhead and communication overhead. Here are our qualitative labels of the computation and communication overheads:

\begin{itemize}[itemsep= 5 pt,topsep = 5 pt, leftmargin =7pt]
	\item Computation overhead: We classify the level of computation overhead into {\em high}, {\em medium}, and {\em low}, with homomorphic encryption, neural network training/inference, and additive/multiplicative noisification as the representative examples, respectively. The high-overhead computation tasks are in general infeasible for IoT end devices. For the medium-overhead computation, IoT end devices are increasingly capable of neural network inference computation due to the emerging inference chips such as Google's Edge TPU \cite{Google'sEdgeTPU}. However, neural network training is still largely infeasible for IoT end devices at present.

	\item Communication overhead: We classify the communication overheads of the existing approaches into three categories: {\em iterative communication}, {\em data swelling}, {\em data retention/compression}, with distributed machine learning, homomorphic encryption, and additive/multiplicative perturbation as the representative examples, respectively. Specifically, distributed machine learning requires iterative model parameter exchanges among the training participants. Such iterative communications will cause significant challenges for IoT networks due to the bandwidth-limited and intermittent communication links. The ciphertexts produced by homomorphic encryption algorithms often have higher data volumes than the plaintexts. In contrast, the additive and multiplicative perturbation will retain and even reduce the data volumes.

\end{itemize}

 {Storage overhead is also a factor of concern for IoT end devices. However, this paper primarily focuses on computation and communication overhead due to the page limit. }

\subsection{Privacy-Preserving Training}
\label{PPLS}

The latest privacy-preserving training approaches that leverage distributed privacy-sensitive data to construct a global ML model or multiple local ML models can be divided into parameter transmission-based (\sect\ref{DML}) and data transmission-based (\sect\ref{Data Transmission-based Approaches}) techniques.

\subsubsection{Parameter Transmission-Based Approaches}

Approaches of this category transmit model parameters instead of data samples for model training. In this way, parameter transmission-based approaches to {privacy-preserving training} push computation towards {participants} rather than the {coordinator}.

\paragraph{Distributed Machine Learning}
\label{DML}
Distributed Machine Learning (DML) is a representative approach of this category.
In DML \cite{brendan2017learning,hamm2015crowd,shokri2015privacy,mcmahan2016communication,dean2012large,zinkevich2010parallelized,boyd2011distributed}, data owners do not reveal their own datasets to anyone in the training phase. In each iteration, the participants upload merely the locally computed parameters or gradients to the coordinator to achieve {\em collaborative learning} \cite{song2018collaborative}. Conventional DML algorithms \cite{zinkevich2010parallelized,hamm2015crowd,shokri2015privacy,boyd2011distributed} exploit the fact that the optimization algorithms based on stochastic gradient descent (SGD) can be parallelized, if the data held by different participants are independently and identically distributed (i.i.d.). {Variants of SGD such as Selective SGD \cite{shokri2015privacy}, parallel SGD \cite{zinkevich2010parallelized}, Alternating Direction Method of Multipliers SGD \cite{boyd2011distributed} and Downpour SGD \cite{dean2012large} are normally used to update the model weights in the distributed fashion.}

{Federated learning \cite{mcmahan2016communication,brendan2017learning} is another prevailing DML approach with more generalized assumptions. In each iteration, a fraction of participants are randomly selected to retrain the model with their local data using the current global model as the starting point and then individually upload the local stochastic gradient descents. The coordinator will then average the gradient descents and update the global model. However, while federated learning \cite{mcmahan2016communication,brendan2017learning} manages to reduce communication overhead, it increases the local computation overhead.}

Arguably, model parameters contain some information about the local training data.
Therefore, in \cite{brendan2017learning,hamm2015crowd,shokri2015privacy}, differential privacy \cite{dwork2011differential} is achieved by adding noises to the locally computed parameter updates. Such schemes thwart definitive inferences about an individual participant if an adversary intentionally collects the obfuscated model updates. Besides, secure data aggregation is applied to aggregate the updates from individual participants \cite{bonawitz2017practical}.  Phong et al. \cite{phong2018privacy} also use additively homomorphic encryption \cite{demillo1978foundations} to encrypt model parameters in the federated learning scheme to prevent information leakage.

\textbf{Limitations:} First, training a deep model locally may be impractical for resource-constrained IoT devices. Second, many iterations are required for the learning process to converge \cite{jiang2019lightweight}, which results in substantial communication overhead. Due to the computation and communication overhead incurred, DML is mainly deployed in the context of cloud computing with enterprise settings. {As shown in \cite{bagdasaryan2018backdoor,bhagoji2018analyzing}, federated learning is vulnerable to backdoor attacks. Hitaj et al. \cite{hitaj2017deep} devise a powerful attack based on generative adversarial networks \cite{goodfellow2014generative} for local data recovery. The attack is still effective even when the communicated parameters are perturbed for differential privacy and secure multi-party computation (MPC) \cite{goldreich1998secure} is applied. }

\paragraph{Model Personalization}
\label{model personalization}
The aim of model personalization \cite{servia2018privacy} is not to learn a global model from privacy-sensitive data owned by the participants, but to learn a personal model for each {participant} based on a public model trained with public data as the starting point. Specifically, the public model is firstly trained with a set of public data at the coordinator and then distributed to each {participant}. Then, each {participant} retrains the model with local data. The idea of transfer leaning \cite{pan2009survey} is leveraged to achieve better performance than the model training with merely local data.

\textbf{Limitations:} This approach may be ineffective for some tasks where the local classes have significantly different patterns from the public data. Additionally, the local training is unsuitable for resource-constrained IoT devices.

\subsubsection{Data Transmission-Based Approaches}
\label{Data Transmission-based Approaches}

 This category of approaches allows {participants} to send local data samples to the honest-but-curious {coordinator}, while protects certain aspect or attribute of the data samples, e.g., user identity, data contents, or raw form of data. It has the following sub-categories: {anonymization}, {cryptographic methods}, {data obfuscation}, and {data synthesis}.

\paragraph{Anonymization}
\label{Anonymization}
Anonymization techniques are designed to anonymize the participant's identity in a group of users, changing the value of quasi-identifiers and removing explicit identifiers. Since the aim is to remove the association between data entries and the data owner, the data samples of interest used for model training remains unchanged. {For field-structured data, anonymization techniques include $k$-anonymity \cite{sweeney2002k}, $l$-diversity \cite{machanavajjhala2006diversity}, and $t$-closeness \cite{li2007t}.} The $k$-same family of algorithms \cite{gross2006model,newton2005preserving,gross2005integrating} are designed to de-identify face images.

 \textbf{Limitations:} As analyzed in \cite{machanavajjhala2006diversity}, anonymization techniques are vulnerable to \textit{homogeneity attack} and \textit{background knowledge attack}. Furthermore, anonymization techniques are traditional data publishing techniques proposed for a centralized database. As commented in a survey \cite{vergara2016privacy}, these anonymization techniques in crowdsensing applications have a main drawback of the need for a trusted proxy to produce the anonymized values and send them to each participant. This implies the risk of single point of failure.

\paragraph{Cryptographic Methods}
\label{Cryptographic Methods}

Cryptographic methods encrypt the training data before transmission. However, traditional cryptographic methods suffer from high computation complexity and the sophistication of key management \cite{eschenauer2002key}. {The method of Homomorphic Encryption (HE) \cite{demillo1978foundations} does not need key propagation and has attracted research interest. With HE, computation on ciphertexts generates an encrypted result which matches the result of the operations performed on the plaintext data after decryption. In \cite{graepel2012ml,zhan2005privacy,qi2008efficient}, the ML model is trained at the coordinator on the HE ciphertexts. During the inference phase, the data is also encrypted before transmission. }

\textbf{Limitations:} However, in HE, operations on the ciphertexts are required to be expressed as polynomials of a bounded degree. Thus, HE is normally applied to the operations with a linear discrimination nature.  Furthermore, HE involves intensive computation and leads to {\em data swelling}. {As benchmarked in \cite{jiang2019lightweight}, HE causes computation overhead millions times higher than a multiplicative obfuscation approach that will be discussed shortly. Additionally, HE will make the training process at least an order of magnitude slower \cite{gilad2016cryptonets}.}

\paragraph{Data Obfuscation} 

\label{Obfuscation}
Data obfuscation methods perturb the data samples used for training a global model. These methods include {additive perturbation}, {multiplicative perturbation}, and {generative obfuscation}.

\begin{itemize}[itemsep= 5 pt,topsep = 5 pt, leftmargin =7pt]
	\item Additive Perturbation: Normally, additive obfuscation is often associated with Differential Privacy (DP) \cite{dwork2011differential}. DP is a formal and quantifiable measure of privacy protection, which can be incorporated in data mining and data publishing \cite{zhu2017differentially}. The key idea of differentially private data mining \cite{abadi2016deep,chaudhuri2009privacy,song2013stochastic} is to learn a model with plaintxt data but perturb the value computed in a certain step (e.g., gradients in optimization) with noises during the training. Such techniques, as discussed earlier, are often used in DML (\sect \ref{DML}). Differentially private data publishing aims to output aggregate information without revealing any specific entry. It can be achieved by adding noises to the query results using Laplacian \cite{dwork2006calibrating}, exponential \cite{mcsherry2007mechanism}, and median \cite{roth2010interactive} mechanisms.

\textbf{Limitations:} {Differentially private data publishing techniques are often used to support the release of limited data representations, such as contingency tables or histograms \cite{zhu2017differentially}. The amount of noise increases dramatically when the queries are correlated \cite{zhu2017differentially}.} Besides, differentially private data publishing often caters into the setting of centralized systems, where a curator collects all the data and respond to queries. But such a trusted curator can be questionable and costly in the context of IoT. Moreover, it incurs the risk of single-point failure. {If the curator is not available, each data contributor perturbs its own result, which leads to an aggregated noise that significantly exceeds the required amount to ensure $\epsilon$-DP of the final result. The experiment in \cite{jiang2019lightweight} shows that, when each participant independently perturbs the training data vectors with Laplacian noise to achieve  $\epsilon$-DP, the support vector macine and deep neural networks learned from the data yield poor accuracy.}

\item {Multiplicative Perturbation}: Random projection \cite{liu2012cloud,jiang2019lightweight,shen2018privacy,chen2005privacy} is a typical multiplicative perturbation. Some random projection schemes \cite{chen2005privacy} preserve the dimensionality of the data but are susceptible to \textit{approximate reconstruction attack} \cite{liu2005random}. Other schemes \cite{liu2012cloud,jiang2019lightweight,shen2018privacy} reduce the dimensions of the data to better preserve privacy. 
{The approach in \cite{shen2018privacy} standardizes the projection matrix $R$ for all participants. However, this design may scale poorly since any collusion between participants would breach data privacy. In \cite{liu2012cloud,jiang2019lightweight}, participants use different private matrices for random projection. Therefore, the Euclidean distances for the perturbed data are no longer preserved, which can significantly degrade the classification accuracy for distance-based classifiers. To tackle this problem, in \cite{liu2012cloud}, the coordinator uses regression to reconstruct the pairwise distances between the original data vectors based on each participant's obfuscated projection results of a set of public data samples. However, the coordinator can use the public samples and their projections to recover random projection matrix of each participant. In \cite{jiang2019lightweight}, rather than using regression for distance estimation, deep neural networks (DNNs) are leveraged to learn the sophisticated pattern of the projected data from multiple participants.}

\textbf{Limitations:} In summary, there exists a trade-off between privacy and utility when applying multiplicative perturbation. If each participant uses a different random projection matrix, privacy can be better preserved but classification accuracy in the cloud is likely degraded.

\item{Generative Obfuscation}: Different from additive/multiplicative perturbation techniques, generative models can also produce obfuscated data. {Huang et al. \cite{huang2017context} propose a Generative Adversarial Privacy (GAP) algorithm, which is composed of a {\em privatizer} and an {\em adversary} network. GAP formulates a minimax game-theoretic problem where the {\em privatizer} aims to obfuscate the original data $X$ to render a specified privacy-sensitive attribute $Y$ non-classifiable by the {\em adversary} network. The {\em privatizer} and {\em adversary} are trained in an iterative manner.}

\textbf{Limitations:} {Running a generative model locally for obfuscation incurs high computation overhead. As a {data publishing} technique, although GAP \cite{huang2017context} restricts the {\em $l_2$} distance between the original data vector and the obfuscated data vector to control the distortion level, there is no guarantee of the utility of the obfuscated data.}

\end{itemize}

\paragraph{Data Synthesis}
\label{Data Synthesis}

Data synthesis methods use generative models that capture the underlying distribution of a private dataset and generate resembling data samples. Such generalization, ideally, would protect individual-specific information. In \cite{acs2018differentially}, differentially private $k$-means clustering is applied on the raw dataset. Then, generative models are trained only on their own cluster using differentially private gradient descent \cite{acs2018differentially}. 

\textbf{Limitations:} Data synthesis is a {data publishing} technique and is usually implemented in a centralized database with massive data samples. As generative models often incur high computation overhead, this approach is not suitable for resource-limited IoT devices.

\subsection{Privacy-Preserving Inference}

\label{PP schemes for Inference}
Compared with a body of research on {privacy-preserving training}, less work is dedicated to {privacy-preserving inference}. {Privacy-preserving inference} approaches assume that the ML model at the coordinator has been previously trained using public plaintext data. 
They aim to protect the privacy contained in test data vectors while maintaining the inference accuracy. Additive perturbation is generally not advisable for deep models because the inference accuracy of deep models can be significantly degraded by small perturbations on input data \cite{zheng2016improving}. In order to achieve privacy preservation in the inference phase against an honest-but-curious coordinator running the ML model, CryptoNets \cite{gilad2016cryptonets} and Multi-party Computation (MPC) \cite{barni2006privacy} are proposed.

\subsubsection{CryptoNets}

Gilad-Bachrach et al. \cite{gilad2016cryptonets} adjust the feed-forward neural network trained with plaintext data so that it can be applied to the homomorphically encrypted data to make encrypted inference. A secret key is needed to decrypt the result. Through the process, not only the data but also the inference result are kept secret against the honest-but-curious coordinator. 

 \textbf{Limitations:} {Unfortunately, the high computational complexity of HE renders CryptoNets unpractical for IoT devices. Moreover, {although CryptoNets does not need to support training over ciphertext,} the neural network still needs to satisfy certain conditions. For example, a square polynomial function instead of a sigmoid or ReLU function should be used as the activation function. However, using square polynomial function as the activation function is rare for existing neural networks. Scaling is also required since encryption scheme does not support floating-point numbers.}

\subsubsection{Multi-Party Computation (MPC)}

MPC \cite{goldreich1998secure} enables the parties involved to jointly compute a function over their inputs while keeping those inputs private. Barni et al. \cite{barni2006privacy} apply MPC in {\em privacy-preserving inference}. Specifically, the participant encrypts the data and sends it to the coordinator. The coordinator computes an inner product between the data and the weights of the first layer and sends the results back to the participant. Then, the participant applies decryption and non-linear transformation. Results are again encrypted before being transmitted to the coordinator. The process continues until all the layers have been computed. In this scheme, the input data and the knowledge embedded in the neural networks are both protected.

\textbf{Limitations}: MPC requires many rounds of communication between the participant and the coordinator, representing considerable communication overhead.

\subsection{Remark}

{The existing approaches reviewed in \sect\ref{PPLS} and \sect\ref{PP schemes for Inference} have different threat, privacy and system models. The anonymization and additive/multiplicative perturbation approaches often introduce affordable overheads and thus are feasible in the context of IoT. However, anonymization mainly focuses on private data publishing and does not address the problem of model training. Thus, additive and multiplicative perturbation approaches are promising for privacy-preserving training in IoT.
In contrast, lightweight privacy-preserving inference approaches for IoT are lacking. Thus, we believe that privacy-preserving inference deserves more research attention.} As such, \sect\ref{sec:ourapproach} describes a lightweight and voluntary privacy-preserving inference approach called {\em ObfNet} and the preliminary results.

\section{ObfNet: Lightweight \& Voluntary Privacy-Preserving Inference}
\label{sec:ourapproach}

\subsection{System, Threat, and Privacy Models}

{\bf System model:} The system consists of multiple resource-constrained participants and a resource-rich coordinator. The coordinator hosts a pre-trained deep model. The participants collect data samples and transmit them to the coordinator for inference using the deep model. The participants do not execute the deep model for inference due to the following reasons. First, each participant has limited computation and storage resources for executing the deep model. Second, in certain scenarios, the deep model may be commercially confidential and should not be released to the participants.

{\bf Threat model:} The threat is an honest-but-curious coordinator. Specifically, the coordinator will not tamper with any data submitted by the participants and the inference results. However, the coordinator is curious about the private information contained in the data submitted by the participants.

{\bf Privacy model:} The raw form of the submitted data is the participant's privacy to be protected. Data form confidentiality is an immediate and basic privacy requirement in many applications.

We discuss several related issues. First, although the inference result may also contain information about the participant, this paper does not consider label privacy. In practice, to mitigate the concern of label privacy leak, the participants can send the inference data samples to the coordinator via anonymity networks, such that the coordinator cannot associate the inference labels with the actual identities of the participants. Second, we aim to design a lightweight and voluntary approach to address the defined privacy threat. It is {\em lightweight} in that the computation performed by a participant should be affordable to resource-constrained IoT objects. {It is {\em voluntary} in that any participant can choose to protect the data privacy by executing {ObfNet} for obfuscation or just send the original data, without needing to inform the coordinator. In other words, the pre-trained deep model at the coordinator admits both original and obfuscated data samples.} This is a desirable feature for legacy deep models.

\begin{figure}
	\includegraphics[width=\linewidth]{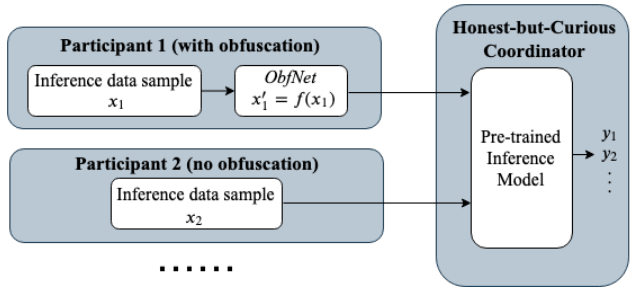}
	\vspace{-1.5em}
	\caption{System model and approach overview.}
	\label{fig:infer}
	\vspace{-1.2em}
	
\end{figure}

\begin{figure}
  \centering
  \begin{subfigure}[b]{0.45\textwidth}
    \includegraphics[width=1\linewidth]{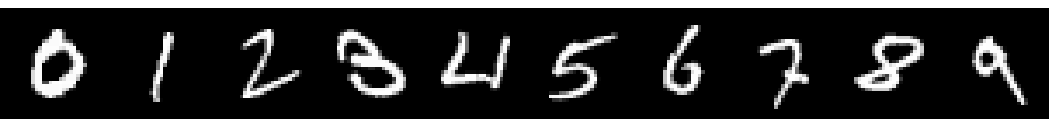}
    \vspace{-1.2em}
    \caption{The original inference data samples.}
    \label{fig:origin}
  \end{subfigure}

  \begin{subfigure}[b]{0.45\textwidth}
    \includegraphics[width=1\linewidth]{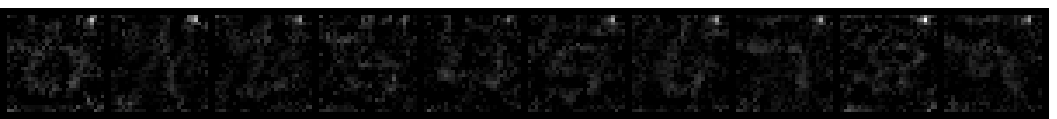}
    \vspace{-1.2em}
    \caption{The data samples obfuscated with ObfNet1.}
    \label{fig:fake1}
  \end{subfigure}

  \begin{subfigure}[b]{0.45\textwidth}
    \includegraphics[width=1\linewidth]{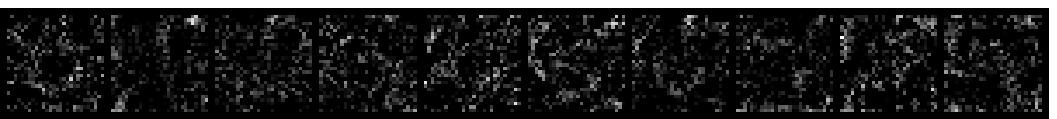}
    \vspace{-1.2em}
    \caption{The data samples obfuscated with ObfNet2-$0$.}
    \label{fig:fake2}
  \end{subfigure}

  \vspace{-0.8em}
  \caption{Inference data samples from MNIST.}
  \label{fig:samples}
  \vspace{-1.2em}
\end{figure}

\subsection{Construction of ObfNet}\label{our_approach}

Fig.~\ref{fig:infer} illustrates our proposed ObfNet approach. The key idea is that we train a small-scale neural network called {\em ObfNet} that gives an output with the same size as the input. The training of ObfNet is as follows. {After the pre-training of the deep inference model,} we use the same training dataset to train the concatenation of the ObfNet and the pre-trained deep inference model. {During the training phase of the concatenated model, only the parameters of the ObfNet are updated while the parameters of the pre-trained inference model are fixed.} As discussed in \sect\ref{sec2}, IoT end devices are in general incapable of ML model training. So, in our approach, the ObfNet is trained by the coordinator and released to the participants for use. {The ObfNet uses many-to-one mapping activation functions, such as the rectified linear unit (ReLU). In that case, the coordinator cannot estimate the exact original data samples based on the obfuscated ones since there exist infinite possible inputs mapping to the same output. For better privacy protection, the coordinator may train multiple ObfNets and send all of them to each participant. The participants can randomly choose one to obfuscate the local data. Moreover, depending on the privacy-sensitivity of the data samples, participants can choose to obfuscate none or part of the data samples. The more ObfNets the coordinator trains and distributes, the less likely the coordinator can figure out which ObfNet is used. However, there exists a trade-off since training and distributing more ObfNets incur more communication overhead.}

\subsection{Evaluation Results with MNIST Dataset}

\begin{figure}
  \includegraphics[width=0.99\linewidth]{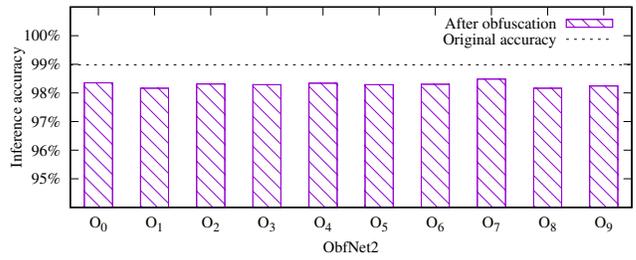}
  \vspace{-0.8em}
  \caption{The inference accuracy of ObfNet2-$x$}
  \label{fig:acc}
  \vspace{-1.2em}

\end{figure}

We evaluate our approach with the MNIST dataset \cite{lecun1998mnist}. The deep inference model is LeNet \cite{lecun1998mnist}, with an inference accuracy of 98.98\%. {We train 11 different ObfNets denoted by ObfNet1 and ObfNet2-$x$, where $x$ is from 0 to 9. ObfNet1 consists of one input layer of $784$ neurons fully connected to an output layer of $784$ neurons with ReLU as the activation function. It has 615,440 learnable parameters with a total volume of 2.17 MB. To show how the obfuscation results change with more hidden layers, we add one more hidden layer of 1,000 neurons in ObfNet2. ObfNet2-$x$, where $x$ is from 0 to 9, are of the same structure and trained individually. Each has 1,569,784 learnable parameters and the total volume of each net is 6.01 MB. The LeNet and ObfNets are implemented using PyTorch \cite{paszke2017automatic}.

Fig.~\ref{fig:samples} shows the original MNIST data samples and the obfuscated samples by ObfNet1 and ObfNet2-$0$. We can see that the obfuscated samples by ObfNet1 still contain some traces of the handwritten digits. However, these traces are not easily recognizable by human inspection. With one more hidden layer, the obfuscated samples by ObfNet2-$0$ are completely unrecognizable. Fig.~\ref{fig:acc} shows the inference accuracy of LeNet when each ObfNet2 is applied to obfuscate the data. As shown in the figure, the difference in the inference accuracy between each ObfNet2 is negligible and the average inference accuracy of the 10 ObfNets is 98.29\%. Compared with the inference accuracy of LeNet on the original data, the accuracy drop is only 0.7\%. Thus, the obfuscation maintains the inference accuracy well.}

\section{Conclusion and Future Work}
\label{sec:con}

This paper reviews the existing privacy-preserving ML approaches that were developed largely in the context of cloud computing and discusses their limitations in the context of IoT. {From our survey, there is a body of research on privacy-preserving training. Additive and multiplicative perturbation approaches are promising for privacy-preserving training in IoT due to their low computation and communication overhead. In contrast, lightweight privacy-preserving inference received limited research. Note that many IoT applications may prefer to use pre-trained deep models. Thus, how to protect the participants' data privacy in using the pre-trained deep models at the coordinator is a meaningful and interesting topic. To this end, we present ObfNet, a lightweight and voluntary privacy-preserving inference approach that obfuscates the data samples while maintaining the inference accuracy of an existing deep neural network.} In our future work, we will perform more extensive evaluation of ObfNet including its performance with other datasets and overhead on actual IoT hardware platforms.

\bibliographystyle{ACM-Reference-Format}
\bibliography{reference}


\begin{thebibliography}{50}

%%% the Web address.

\ifx \showCODEN    \undefined \def \showCODEN     #1{\unskip}     \fi
\ifx \showDOI      \undefined \def \showDOI       #1{#1}\fi
\ifx \showISBNx    \undefined \def \showISBNx     #1{\unskip}     \fi
\ifx \showISBNxiii \undefined \def \showISBNxiii  #1{\unskip}     \fi
\ifx \showISSN     \undefined \def \showISSN      #1{\unskip}     \fi
\ifx \showLCCN     \undefined \def \showLCCN      #1{\unskip}     \fi
\ifx \shownote     \undefined \def \shownote      #1{#1}          \fi
\ifx \showarticletitle \undefined \def \showarticletitle #1{#1}   \fi
\ifx \showURL      \undefined \def \showURL       {\relax}        \fi
% invisible to TeX
\providecommand\bibfield[2]{#2}
\providecommand\bibinfo[2]{#2}
\providecommand\natexlab[1]{#1}
\providecommand\showeprint[2][]{arXiv:#2}

\bibitem[\protect\citeauthoryear{Abadi, Chu, Goodfellow, McMahan, Mironov,
  Talwar, and Zhang}{Abadi et~al\mbox{.}}{2016}]%
        {abadi2016deep}
\bibfield{author}{\bibinfo{person}{Martin Abadi}, \bibinfo{person}{Andy Chu},
  \bibinfo{person}{Ian Goodfellow}, \bibinfo{person}{H~Brendan McMahan},
  \bibinfo{person}{Ilya Mironov}, \bibinfo{person}{Kunal Talwar}, {and}
  \bibinfo{person}{Li Zhang}.} \bibinfo{year}{2016}\natexlab{}.
\newblock \showarticletitle{Deep learning with differential privacy}. In
  \bibinfo{booktitle}{\emph{Proceedings of the 2016 ACM SIGSAC Conference on
  Computer and Communications Security}}. ACM, \bibinfo{pages}{308--318}.
\newblock


\bibitem[\protect\citeauthoryear{Acs, Melis, Castelluccia, and
  De~Cristofaro}{Acs et~al\mbox{.}}{2018}]%
        {acs2018differentially}
\bibfield{author}{\bibinfo{person}{Gergely Acs}, \bibinfo{person}{Luca Melis},
  \bibinfo{person}{Claude Castelluccia}, {and} \bibinfo{person}{Emiliano
  De~Cristofaro}.} \bibinfo{year}{2018}\natexlab{}.
\newblock \showarticletitle{Differentially private mixture of generative neural
  networks}.
\newblock \bibinfo{journal}{\emph{IEEE Transactions on Knowledge and Data
  Engineering}} \bibinfo{volume}{31}, \bibinfo{number}{6}
  (\bibinfo{year}{2018}), \bibinfo{pages}{1109--1121}.
\newblock


\bibitem[\protect\citeauthoryear{Bagdasaryan, Veit, Hua, Estrin, and
  Shmatikov}{Bagdasaryan et~al\mbox{.}}{2018}]%
        {bagdasaryan2018backdoor}
\bibfield{author}{\bibinfo{person}{Eugene Bagdasaryan},
  \bibinfo{person}{Andreas Veit}, \bibinfo{person}{Yiqing Hua},
  \bibinfo{person}{Deborah Estrin}, {and} \bibinfo{person}{Vitaly Shmatikov}.}
  \bibinfo{year}{2018}\natexlab{}.
\newblock \showarticletitle{How to backdoor federated learning}.
\newblock \bibinfo{journal}{\emph{arXiv preprint arXiv:1807.00459}}
  (\bibinfo{year}{2018}).
\newblock


\bibitem[\protect\citeauthoryear{Barni, Orlandi, and Piva}{Barni
  et~al\mbox{.}}{2006}]%
        {barni2006privacy}
\bibfield{author}{\bibinfo{person}{Mauro Barni}, \bibinfo{person}{Claudio
  Orlandi}, {and} \bibinfo{person}{Alessandro Piva}.}
  \bibinfo{year}{2006}\natexlab{}.
\newblock \showarticletitle{A privacy-preserving protocol for
  neural-network-based computation}. In \bibinfo{booktitle}{\emph{Proceedings
  of the 8th workshop on Multimedia and security}}. ACM,
  \bibinfo{pages}{146--151}.
\newblock


\bibitem[\protect\citeauthoryear{Bhagoji, Chakraborty, Mittal, and
  Calo}{Bhagoji et~al\mbox{.}}{2018}]%
        {bhagoji2018analyzing}
\bibfield{author}{\bibinfo{person}{Arjun~Nitin Bhagoji},
  \bibinfo{person}{Supriyo Chakraborty}, \bibinfo{person}{Prateek Mittal},
  {and} \bibinfo{person}{Seraphin Calo}.} \bibinfo{year}{2018}\natexlab{}.
\newblock \showarticletitle{Analyzing federated learning through an adversarial
  lens}.
\newblock \bibinfo{journal}{\emph{arXiv preprint arXiv:1811.12470}}
  (\bibinfo{year}{2018}).
\newblock


\bibitem[\protect\citeauthoryear{Bonawitz, Ivanov, Kreuter, Marcedone, McMahan,
  Patel, Ramage, Segal, and Seth}{Bonawitz et~al\mbox{.}}{2017}]%
        {bonawitz2017practical}
\bibfield{author}{\bibinfo{person}{Keith Bonawitz}, \bibinfo{person}{Vladimir
  Ivanov}, \bibinfo{person}{Ben Kreuter}, \bibinfo{person}{Antonio Marcedone},
  \bibinfo{person}{H~Brendan McMahan}, \bibinfo{person}{Sarvar Patel},
  \bibinfo{person}{Daniel Ramage}, \bibinfo{person}{Aaron Segal}, {and}
  \bibinfo{person}{Karn Seth}.} \bibinfo{year}{2017}\natexlab{}.
\newblock \showarticletitle{Practical secure aggregation for privacy-preserving
  machine learning}. In \bibinfo{booktitle}{\emph{Proceedings of the 2017 ACM
  SIGSAC Conference on Computer and Communications Security}}. ACM,
  \bibinfo{pages}{1175--1191}.
\newblock


\bibitem[\protect\citeauthoryear{Boyd, Parikh, Chu, Peleato, Eckstein,
  et~al\mbox{.}}{Boyd et~al\mbox{.}}{2011}]%
        {boyd2011distributed}
\bibfield{author}{\bibinfo{person}{Stephen Boyd}, \bibinfo{person}{Neal
  Parikh}, \bibinfo{person}{Eric Chu}, \bibinfo{person}{Borja Peleato},
  \bibinfo{person}{Jonathan Eckstein}, {et~al\mbox{.}}}
  \bibinfo{year}{2011}\natexlab{}.
\newblock \showarticletitle{Distributed optimization and statistical learning
  via the alternating direction method of multipliers}.
\newblock \bibinfo{journal}{\emph{Foundations and Trends{\textregistered} in
  Machine learning}} \bibinfo{volume}{3}, \bibinfo{number}{1}
  (\bibinfo{year}{2011}), \bibinfo{pages}{1--122}.
\newblock


\bibitem[\protect\citeauthoryear{Brendan~McMahan, Ramage, Talwar, and
  Zhang}{Brendan~McMahan et~al\mbox{.}}{2017}]%
        {brendan2017learning}
\bibfield{author}{\bibinfo{person}{H Brendan~McMahan}, \bibinfo{person}{Daniel
  Ramage}, \bibinfo{person}{Kunal Talwar}, {and} \bibinfo{person}{Li Zhang}.}
  \bibinfo{year}{2017}\natexlab{}.
\newblock \showarticletitle{Learning Differentially Private Recurrent Language
  Models}.
\newblock \bibinfo{journal}{\emph{arXiv preprint arXiv:1710.06963}}
  (\bibinfo{year}{2017}).
\newblock


\bibitem[\protect\citeauthoryear{Chaudhuri and Monteleoni}{Chaudhuri and
  Monteleoni}{2009}]%
        {chaudhuri2009privacy}
\bibfield{author}{\bibinfo{person}{Kamalika Chaudhuri} {and}
  \bibinfo{person}{Claire Monteleoni}.} \bibinfo{year}{2009}\natexlab{}.
\newblock \showarticletitle{Privacy-preserving logistic regression}. In
  \bibinfo{booktitle}{\emph{Advances in neural information processing
  systems}}. \bibinfo{pages}{289--296}.
\newblock


\bibitem[\protect\citeauthoryear{Chen and Liu}{Chen and Liu}{2005}]%
        {chen2005privacy}
\bibfield{author}{\bibinfo{person}{Keke Chen} {and} \bibinfo{person}{Ling
  Liu}.} \bibinfo{year}{2005}\natexlab{}.
\newblock \showarticletitle{Privacy preserving data classification with
  rotation perturbation}. In \bibinfo{booktitle}{\emph{Fifth IEEE International
  Conference on Data Mining (ICDM'05)}}. IEEE, \bibinfo{pages}{4--pp}.
\newblock


\bibitem[\protect\citeauthoryear{Dean, Corrado, Monga, Chen, Devin, Mao,
  Senior, Tucker, Yang, Le, et~al\mbox{.}}{Dean et~al\mbox{.}}{2012}]%
        {dean2012large}
\bibfield{author}{\bibinfo{person}{Jeffrey Dean}, \bibinfo{person}{Greg
  Corrado}, \bibinfo{person}{Rajat Monga}, \bibinfo{person}{Kai Chen},
  \bibinfo{person}{Matthieu Devin}, \bibinfo{person}{Mark Mao},
  \bibinfo{person}{Andrew Senior}, \bibinfo{person}{Paul Tucker},
  \bibinfo{person}{Ke Yang}, \bibinfo{person}{Quoc~V Le}, {et~al\mbox{.}}}
  \bibinfo{year}{2012}\natexlab{}.
\newblock \showarticletitle{Large scale distributed deep networks}. In
  \bibinfo{booktitle}{\emph{Advances in neural information processing
  systems}}. \bibinfo{pages}{1223--1231}.
\newblock


\bibitem[\protect\citeauthoryear{DeMillo}{DeMillo}{1978}]%
        {demillo1978foundations}
\bibfield{author}{\bibinfo{person}{Richard~Alan DeMillo}.}
  \bibinfo{year}{1978}\natexlab{}.
\newblock \bibinfo{booktitle}{\emph{Foundations of secure computation}}.
\newblock \bibinfo{type}{{T}echnical {R}eport}. \bibinfo{institution}{Georgia
  Institute of Technology}.
\newblock


\bibitem[\protect\citeauthoryear{Dwork}{Dwork}{2011}]%
        {dwork2011differential}
\bibfield{author}{\bibinfo{person}{Cynthia Dwork}.}
  \bibinfo{year}{2011}\natexlab{}.
\newblock \showarticletitle{Differential privacy}.
\newblock \bibinfo{journal}{\emph{Encyclopedia of Cryptography and Security}}
  (\bibinfo{year}{2011}), \bibinfo{pages}{338--340}.
\newblock


\bibitem[\protect\citeauthoryear{Dwork, McSherry, Nissim, and Smith}{Dwork
  et~al\mbox{.}}{2006}]%
        {dwork2006calibrating}
\bibfield{author}{\bibinfo{person}{Cynthia Dwork}, \bibinfo{person}{Frank
  McSherry}, \bibinfo{person}{Kobbi Nissim}, {and} \bibinfo{person}{Adam
  Smith}.} \bibinfo{year}{2006}\natexlab{}.
\newblock \showarticletitle{Calibrating noise to sensitivity in private data
  analysis}. In \bibinfo{booktitle}{\emph{Theory of cryptography conference}}.
  Springer, \bibinfo{pages}{265--284}.
\newblock


\bibitem[\protect\citeauthoryear{Eschenauer and Gligor}{Eschenauer and
  Gligor}{2002}]%
        {eschenauer2002key}
\bibfield{author}{\bibinfo{person}{Laurent Eschenauer} {and}
  \bibinfo{person}{Virgil~D Gligor}.} \bibinfo{year}{2002}\natexlab{}.
\newblock \showarticletitle{A key-management scheme for distributed sensor
  networks}. In \bibinfo{booktitle}{\emph{Proceedings of the 9th ACM conference
  on Computer and communications security}}. ACM, \bibinfo{pages}{41--47}.
\newblock


\bibitem[\protect\citeauthoryear{Gilad-Bachrach, Dowlin, Laine, Lauter,
  Naehrig, and Wernsing}{Gilad-Bachrach et~al\mbox{.}}{2016}]%
        {gilad2016cryptonets}
\bibfield{author}{\bibinfo{person}{Ran Gilad-Bachrach}, \bibinfo{person}{Nathan
  Dowlin}, \bibinfo{person}{Kim Laine}, \bibinfo{person}{Kristin Lauter},
  \bibinfo{person}{Michael Naehrig}, {and} \bibinfo{person}{John Wernsing}.}
  \bibinfo{year}{2016}\natexlab{}.
\newblock \showarticletitle{Cryptonets: Applying neural networks to encrypted
  data with high throughput and accuracy}. In
  \bibinfo{booktitle}{\emph{International Conference on Machine Learning}}.
  \bibinfo{pages}{201--210}.
\newblock


\bibitem[\protect\citeauthoryear{Goldreich}{Goldreich}{1998}]%
        {goldreich1998secure}
\bibfield{author}{\bibinfo{person}{Oded Goldreich}.}
  \bibinfo{year}{1998}\natexlab{}.
\newblock \showarticletitle{Secure multi-party computation}.
\newblock \bibinfo{journal}{\emph{Manuscript. Preliminary version}}
  \bibinfo{volume}{78} (\bibinfo{year}{1998}).
\newblock


\bibitem[\protect\citeauthoryear{Goodfellow, Pouget-Abadie, Mirza, Xu,
  Warde-Farley, Ozair, Courville, and Bengio}{Goodfellow et~al\mbox{.}}{2014}]%
        {goodfellow2014generative}
\bibfield{author}{\bibinfo{person}{Ian Goodfellow}, \bibinfo{person}{Jean
  Pouget-Abadie}, \bibinfo{person}{Mehdi Mirza}, \bibinfo{person}{Bing Xu},
  \bibinfo{person}{David Warde-Farley}, \bibinfo{person}{Sherjil Ozair},
  \bibinfo{person}{Aaron Courville}, {and} \bibinfo{person}{Yoshua Bengio}.}
  \bibinfo{year}{2014}\natexlab{}.
\newblock \showarticletitle{Generative adversarial nets}. In
  \bibinfo{booktitle}{\emph{Advances in neural information processing
  systems}}. \bibinfo{pages}{2672--2680}.
\newblock


\bibitem[\protect\citeauthoryear{Google}{Google}{[n.d.]}]%
        {Google'sEdgeTPU}
\bibfield{author}{\bibinfo{person}{Google}.} \bibinfo{year}{[n.d.]}\natexlab{}.
\newblock \bibinfo{title}{Google's Edge TPU}.
\newblock \bibinfo{howpublished}{\url{https://cloud.google.com/edge-tpu/}}.
\newblock


\bibitem[\protect\citeauthoryear{Graepel, Lauter, and Naehrig}{Graepel
  et~al\mbox{.}}{2012}]%
        {graepel2012ml}
\bibfield{author}{\bibinfo{person}{Thore Graepel}, \bibinfo{person}{Kristin
  Lauter}, {and} \bibinfo{person}{Michael Naehrig}.}
  \bibinfo{year}{2012}\natexlab{}.
\newblock \showarticletitle{ML confidential: Machine learning on encrypted
  data}. In \bibinfo{booktitle}{\emph{International Conference on Information
  Security and Cryptology}}. Springer, \bibinfo{pages}{1--21}.
\newblock


\bibitem[\protect\citeauthoryear{Gross, Airoldi, Malin, and Sweeney}{Gross
  et~al\mbox{.}}{2005}]%
        {gross2005integrating}
\bibfield{author}{\bibinfo{person}{Ralph Gross}, \bibinfo{person}{Edoardo
  Airoldi}, \bibinfo{person}{Bradley Malin}, {and} \bibinfo{person}{Latanya
  Sweeney}.} \bibinfo{year}{2005}\natexlab{}.
\newblock \showarticletitle{Integrating utility into face de-identification}.
  In \bibinfo{booktitle}{\emph{International Workshop on Privacy Enhancing
  Technologies}}. Springer, \bibinfo{pages}{227--242}.
\newblock


\bibitem[\protect\citeauthoryear{Gross, Sweeney, De~la Torre, and Baker}{Gross
  et~al\mbox{.}}{2006}]%
        {gross2006model}
\bibfield{author}{\bibinfo{person}{Ralph Gross}, \bibinfo{person}{Latanya
  Sweeney}, \bibinfo{person}{Fernando De~la Torre}, {and}
  \bibinfo{person}{Simon Baker}.} \bibinfo{year}{2006}\natexlab{}.
\newblock \showarticletitle{Model-based face de-identification}. In
  \bibinfo{booktitle}{\emph{2006 Conference on Computer Vision and Pattern
  Recognition Workshop (CVPRW'06)}}. IEEE, \bibinfo{pages}{161--161}.
\newblock


\bibitem[\protect\citeauthoryear{Hamm, Champion, Chen, Belkin, and Xuan}{Hamm
  et~al\mbox{.}}{2015}]%
        {hamm2015crowd}
\bibfield{author}{\bibinfo{person}{Jihun Hamm}, \bibinfo{person}{Adam~C
  Champion}, \bibinfo{person}{Guoxing Chen}, \bibinfo{person}{Mikhail Belkin},
  {and} \bibinfo{person}{Dong Xuan}.} \bibinfo{year}{2015}\natexlab{}.
\newblock \showarticletitle{Crowd-ml: A privacy-preserving learning framework
  for a crowd of smart devices}. In \bibinfo{booktitle}{\emph{2015 IEEE 35th
  International Conference on Distributed Computing Systems}}. IEEE,
  \bibinfo{pages}{11--20}.
\newblock


\bibitem[\protect\citeauthoryear{Hitaj, Ateniese, and Perez-Cruz}{Hitaj
  et~al\mbox{.}}{2017}]%
        {hitaj2017deep}
\bibfield{author}{\bibinfo{person}{Briland Hitaj}, \bibinfo{person}{Giuseppe
  Ateniese}, {and} \bibinfo{person}{Fernando Perez-Cruz}.}
  \bibinfo{year}{2017}\natexlab{}.
\newblock \showarticletitle{Deep models under the GAN: information leakage from
  collaborative deep learning}. In \bibinfo{booktitle}{\emph{Proceedings of the
  2017 ACM SIGSAC Conference on Computer and Communications Security}}. ACM,
  \bibinfo{pages}{603--618}.
\newblock


\bibitem[\protect\citeauthoryear{Huang, Kairouz, Chen, Sankar, and
  Rajagopal}{Huang et~al\mbox{.}}{2017}]%
        {huang2017context}
\bibfield{author}{\bibinfo{person}{Chong Huang}, \bibinfo{person}{Peter
  Kairouz}, \bibinfo{person}{Xiao Chen}, \bibinfo{person}{Lalitha Sankar},
  {and} \bibinfo{person}{Ram Rajagopal}.} \bibinfo{year}{2017}\natexlab{}.
\newblock \showarticletitle{Context-aware generative adversarial privacy}.
\newblock \bibinfo{journal}{\emph{Entropy}} \bibinfo{volume}{19},
  \bibinfo{number}{12} (\bibinfo{year}{2017}), \bibinfo{pages}{656}.
\newblock


\bibitem[\protect\citeauthoryear{Jiang, Tan, Lou, and Lin}{Jiang
  et~al\mbox{.}}{2019}]%
        {jiang2019lightweight}
\bibfield{author}{\bibinfo{person}{Linshan Jiang}, \bibinfo{person}{Rui Tan},
  \bibinfo{person}{Xin Lou}, {and} \bibinfo{person}{Guosheng Lin}.}
  \bibinfo{year}{2019}\natexlab{}.
\newblock \showarticletitle{On Lightweight Privacy-Preserving Collaborative
  Learning for internet-of-things objects}.
\newblock  (\bibinfo{year}{2019}), \bibinfo{pages}{70--81}.
\newblock


\bibitem[\protect\citeauthoryear{LeCun}{LeCun}{[n.d.]}]%
        {lecun1998mnist}
\bibfield{author}{\bibinfo{person}{Yann LeCun}.}
  \bibinfo{year}{[n.d.]}\natexlab{}.
\newblock \showarticletitle{The MNIST database of handwritten digits}.
\newblock \bibinfo{journal}{\emph{http://yann. lecun. com/exdb/mnist/}}
  (\bibinfo{year}{[n.\,d.]}).
\newblock


\bibitem[\protect\citeauthoryear{Li, Li, and Venkatasubramanian}{Li
  et~al\mbox{.}}{2007}]%
        {li2007t}
\bibfield{author}{\bibinfo{person}{Ninghui Li}, \bibinfo{person}{Tiancheng Li},
  {and} \bibinfo{person}{Suresh Venkatasubramanian}.}
  \bibinfo{year}{2007}\natexlab{}.
\newblock \showarticletitle{t-closeness: Privacy beyond k-anonymity and
  l-diversity}. In \bibinfo{booktitle}{\emph{2007 IEEE 23rd International
  Conference on Data Engineering}}. IEEE, \bibinfo{pages}{106--115}.
\newblock


\bibitem[\protect\citeauthoryear{Liu, Jiang, Sha, and Govindan}{Liu
  et~al\mbox{.}}{2012}]%
        {liu2012cloud}
\bibfield{author}{\bibinfo{person}{Bin Liu}, \bibinfo{person}{Yurong Jiang},
  \bibinfo{person}{Fei Sha}, {and} \bibinfo{person}{Ramesh Govindan}.}
  \bibinfo{year}{2012}\natexlab{}.
\newblock \showarticletitle{Cloud-enabled privacy-preserving collaborative
  learning for mobile sensing}. In \bibinfo{booktitle}{\emph{Proceedings of the
  10th ACM Conference on Embedded Network Sensor Systems}}. ACM,
  \bibinfo{pages}{57--70}.
\newblock


\bibitem[\protect\citeauthoryear{Liu, Kargupta, and Ryan}{Liu
  et~al\mbox{.}}{2005}]%
        {liu2005random}
\bibfield{author}{\bibinfo{person}{Kun Liu}, \bibinfo{person}{Hillol Kargupta},
  {and} \bibinfo{person}{Jessica Ryan}.} \bibinfo{year}{2005}\natexlab{}.
\newblock \showarticletitle{Random projection-based multiplicative data
  perturbation for privacy preserving distributed data mining}.
\newblock \bibinfo{journal}{\emph{IEEE Transactions on knowledge and Data
  Engineering}} \bibinfo{volume}{18}, \bibinfo{number}{1}
  (\bibinfo{year}{2005}), \bibinfo{pages}{92--106}.
\newblock


\bibitem[\protect\citeauthoryear{Machanavajjhala, Gehrke, Kifer, and
  Venkitasubramaniam}{Machanavajjhala et~al\mbox{.}}{2006}]%
        {machanavajjhala2006diversity}
\bibfield{author}{\bibinfo{person}{Ashwin Machanavajjhala},
  \bibinfo{person}{Johannes Gehrke}, \bibinfo{person}{Daniel Kifer}, {and}
  \bibinfo{person}{Muthuramakrishnan Venkitasubramaniam}.}
  \bibinfo{year}{2006}\natexlab{}.
\newblock \showarticletitle{l-diversity: Privacy beyond k-anonymity}. In
  \bibinfo{booktitle}{\emph{22nd International Conference on Data Engineering
  (ICDE'06)}}. IEEE, \bibinfo{pages}{24--24}.
\newblock


\bibitem[\protect\citeauthoryear{McMahan, Moore, Ramage, Hampson,
  et~al\mbox{.}}{McMahan et~al\mbox{.}}{2016}]%
        {mcmahan2016communication}
\bibfield{author}{\bibinfo{person}{H~Brendan McMahan}, \bibinfo{person}{Eider
  Moore}, \bibinfo{person}{Daniel Ramage}, \bibinfo{person}{Seth Hampson},
  {et~al\mbox{.}}} \bibinfo{year}{2016}\natexlab{}.
\newblock \showarticletitle{Communication-efficient learning of deep networks
  from decentralized data}.
\newblock \bibinfo{journal}{\emph{arXiv preprint arXiv:1602.05629}}
  (\bibinfo{year}{2016}).
\newblock


\bibitem[\protect\citeauthoryear{McSherry and Talwar}{McSherry and
  Talwar}{2007}]%
        {mcsherry2007mechanism}
\bibfield{author}{\bibinfo{person}{Frank McSherry} {and} \bibinfo{person}{Kunal
  Talwar}.} \bibinfo{year}{2007}\natexlab{}.
\newblock \showarticletitle{Mechanism Design via Differential Privacy.}. In
  \bibinfo{booktitle}{\emph{FOCS}}, Vol.~\bibinfo{volume}{7}.
  \bibinfo{pages}{94--103}.
\newblock


\bibitem[\protect\citeauthoryear{Newton, Sweeney, and Malin}{Newton
  et~al\mbox{.}}{2005}]%
        {newton2005preserving}
\bibfield{author}{\bibinfo{person}{Elaine~M Newton}, \bibinfo{person}{Latanya
  Sweeney}, {and} \bibinfo{person}{Bradley Malin}.}
  \bibinfo{year}{2005}\natexlab{}.
\newblock \showarticletitle{Preserving privacy by de-identifying face images}.
\newblock \bibinfo{journal}{\emph{IEEE transactions on Knowledge and Data
  Engineering}} \bibinfo{volume}{17}, \bibinfo{number}{2}
  (\bibinfo{year}{2005}), \bibinfo{pages}{232--243}.
\newblock


\bibitem[\protect\citeauthoryear{Pan and Yang}{Pan and Yang}{2009}]%
        {pan2009survey}
\bibfield{author}{\bibinfo{person}{Sinno~Jialin Pan} {and}
  \bibinfo{person}{Qiang Yang}.} \bibinfo{year}{2009}\natexlab{}.
\newblock \showarticletitle{A survey on transfer learning}.
\newblock \bibinfo{journal}{\emph{IEEE Transactions on knowledge and data
  engineering}} \bibinfo{volume}{22}, \bibinfo{number}{10}
  (\bibinfo{year}{2009}), \bibinfo{pages}{1345--1359}.
\newblock


\bibitem[\protect\citeauthoryear{Paszke, Gross, Chintala, Chanan, Yang, DeVito,
  Lin, Desmaison, Antiga, and Lerer}{Paszke et~al\mbox{.}}{2017}]%
        {paszke2017automatic}
\bibfield{author}{\bibinfo{person}{Adam Paszke}, \bibinfo{person}{Sam Gross},
  \bibinfo{person}{Soumith Chintala}, \bibinfo{person}{Gregory Chanan},
  \bibinfo{person}{Edward Yang}, \bibinfo{person}{Zachary DeVito},
  \bibinfo{person}{Zeming Lin}, \bibinfo{person}{Alban Desmaison},
  \bibinfo{person}{Luca Antiga}, {and} \bibinfo{person}{Adam Lerer}.}
  \bibinfo{year}{2017}\natexlab{}.
\newblock \showarticletitle{Automatic Differentiation in {PyTorch}}. In
  \bibinfo{booktitle}{\emph{NIPS Autodiff Workshop}}.
\newblock


\bibitem[\protect\citeauthoryear{Phong, Aono, Hayashi, Wang, and Moriai}{Phong
  et~al\mbox{.}}{2018}]%
        {phong2018privacy}
\bibfield{author}{\bibinfo{person}{Le~Trieu Phong}, \bibinfo{person}{Yoshinori
  Aono}, \bibinfo{person}{Takuya Hayashi}, \bibinfo{person}{Lihua Wang}, {and}
  \bibinfo{person}{Shiho Moriai}.} \bibinfo{year}{2018}\natexlab{}.
\newblock \showarticletitle{Privacy-preserving deep learning via additively
  homomorphic encryption}.
\newblock \bibinfo{journal}{\emph{IEEE Transactions on Information Forensics
  and Security}} \bibinfo{volume}{13}, \bibinfo{number}{5}
  (\bibinfo{year}{2018}), \bibinfo{pages}{1333--1345}.
\newblock


\bibitem[\protect\citeauthoryear{Qi and Atallah}{Qi and Atallah}{2008}]%
        {qi2008efficient}
\bibfield{author}{\bibinfo{person}{Yinian Qi} {and} \bibinfo{person}{Mikhail~J
  Atallah}.} \bibinfo{year}{2008}\natexlab{}.
\newblock \showarticletitle{Efficient privacy-preserving k-nearest neighbor
  search}. In \bibinfo{booktitle}{\emph{2008 The 28th International Conference
  on Distributed Computing Systems}}. IEEE, \bibinfo{pages}{311--319}.
\newblock


\bibitem[\protect\citeauthoryear{Roth and Roughgarden}{Roth and
  Roughgarden}{2010}]%
        {roth2010interactive}
\bibfield{author}{\bibinfo{person}{Aaron Roth} {and} \bibinfo{person}{Tim
  Roughgarden}.} \bibinfo{year}{2010}\natexlab{}.
\newblock \showarticletitle{Interactive privacy via the median mechanism}. In
  \bibinfo{booktitle}{\emph{Proceedings of the forty-second ACM symposium on
  Theory of computing}}. ACM, \bibinfo{pages}{765--774}.
\newblock


\bibitem[\protect\citeauthoryear{Servia-Rodr{\'\i}guez, Wang, Zhao, Mortier,
  and Haddadi}{Servia-Rodr{\'\i}guez et~al\mbox{.}}{2018}]%
        {servia2018privacy}
\bibfield{author}{\bibinfo{person}{Sandra Servia-Rodr{\'\i}guez},
  \bibinfo{person}{Liang Wang}, \bibinfo{person}{Jianxin~R Zhao},
  \bibinfo{person}{Richard Mortier}, {and} \bibinfo{person}{Hamed Haddadi}.}
  \bibinfo{year}{2018}\natexlab{}.
\newblock \showarticletitle{Privacy-Preserving Personal Model Training}. In
  \bibinfo{booktitle}{\emph{2018 IEEE/ACM Third International Conference on
  Internet-of-Things Design and Implementation (IoTDI)}}. IEEE,
  \bibinfo{pages}{153--164}.
\newblock


\bibitem[\protect\citeauthoryear{Shen, Luo, Yin, Wen, Daniela, and Hu}{Shen
  et~al\mbox{.}}{2018}]%
        {shen2018privacy}
\bibfield{author}{\bibinfo{person}{Yiran Shen}, \bibinfo{person}{Chengwen Luo},
  \bibinfo{person}{Dan Yin}, \bibinfo{person}{Hongkai Wen},
  \bibinfo{person}{Rus Daniela}, {and} \bibinfo{person}{Wen Hu}.}
  \bibinfo{year}{2018}\natexlab{}.
\newblock \showarticletitle{Privacy-preserving sparse representation
  classification in cloud-enabled mobile applications}.
\newblock \bibinfo{journal}{\emph{Computer Networks}}  \bibinfo{volume}{133}
  (\bibinfo{year}{2018}), \bibinfo{pages}{59--72}.
\newblock


\bibitem[\protect\citeauthoryear{Shokri and Shmatikov}{Shokri and
  Shmatikov}{2015}]%
        {shokri2015privacy}
\bibfield{author}{\bibinfo{person}{Reza Shokri} {and} \bibinfo{person}{Vitaly
  Shmatikov}.} \bibinfo{year}{2015}\natexlab{}.
\newblock \showarticletitle{Privacy-preserving deep learning}. In
  \bibinfo{booktitle}{\emph{Proceedings of the 22nd ACM SIGSAC conference on
  computer and communications security}}. ACM, \bibinfo{pages}{1310--1321}.
\newblock


\bibitem[\protect\citeauthoryear{Song and Chai}{Song and Chai}{2018}]%
        {song2018collaborative}
\bibfield{author}{\bibinfo{person}{Guocong Song} {and} \bibinfo{person}{Wei
  Chai}.} \bibinfo{year}{2018}\natexlab{}.
\newblock \showarticletitle{Collaborative learning for deep neural networks}.
  In \bibinfo{booktitle}{\emph{Advances in Neural Information Processing
  Systems}}. \bibinfo{pages}{1832--1841}.
\newblock


\bibitem[\protect\citeauthoryear{Song, Chaudhuri, and Sarwate}{Song
  et~al\mbox{.}}{2013}]%
        {song2013stochastic}
\bibfield{author}{\bibinfo{person}{Shuang Song}, \bibinfo{person}{Kamalika
  Chaudhuri}, {and} \bibinfo{person}{Anand~D Sarwate}.}
  \bibinfo{year}{2013}\natexlab{}.
\newblock \showarticletitle{Stochastic gradient descent with differentially
  private updates}. In \bibinfo{booktitle}{\emph{2013 IEEE Global Conference on
  Signal and Information Processing}}. IEEE, \bibinfo{pages}{245--248}.
\newblock


\bibitem[\protect\citeauthoryear{Sweeney}{Sweeney}{2002}]%
        {sweeney2002k}
\bibfield{author}{\bibinfo{person}{Latanya Sweeney}.}
  \bibinfo{year}{2002}\natexlab{}.
\newblock \showarticletitle{k-anonymity: A model for protecting privacy}.
\newblock \bibinfo{journal}{\emph{International Journal of Uncertainty,
  Fuzziness and Knowledge-Based Systems}} \bibinfo{volume}{10},
  \bibinfo{number}{05} (\bibinfo{year}{2002}), \bibinfo{pages}{557--570}.
\newblock


\bibitem[\protect\citeauthoryear{Vergara-Laurens, Jaimes, and
  Labrador}{Vergara-Laurens et~al\mbox{.}}{2016}]%
        {vergara2016privacy}
\bibfield{author}{\bibinfo{person}{Idalides~J Vergara-Laurens},
  \bibinfo{person}{Luis~G Jaimes}, {and} \bibinfo{person}{Miguel~A Labrador}.}
  \bibinfo{year}{2016}\natexlab{}.
\newblock \showarticletitle{Privacy-preserving mechanisms for crowdsensing:
  Survey and research challenges}.
\newblock \bibinfo{journal}{\emph{IEEE Internet of Things Journal}}
  \bibinfo{volume}{4}, \bibinfo{number}{4} (\bibinfo{year}{2016}),
  \bibinfo{pages}{855--869}.
\newblock


\bibitem[\protect\citeauthoryear{Zhan, Chang, and Matwin}{Zhan
  et~al\mbox{.}}{2005}]%
        {zhan2005privacy}
\bibfield{author}{\bibinfo{person}{Justin~Zhijun Zhan}, \bibinfo{person}{LiWu
  Chang}, {and} \bibinfo{person}{Stan Matwin}.}
  \bibinfo{year}{2005}\natexlab{}.
\newblock \showarticletitle{Privacy preserving k-nearest neighbor
  classification.}
\newblock \bibinfo{journal}{\emph{IJ Network Security}} \bibinfo{volume}{1},
  \bibinfo{number}{1} (\bibinfo{year}{2005}), \bibinfo{pages}{46--51}.
\newblock


\bibitem[\protect\citeauthoryear{Zheng, Song, Leung, and Goodfellow}{Zheng
  et~al\mbox{.}}{2016}]%
        {zheng2016improving}
\bibfield{author}{\bibinfo{person}{Stephan Zheng}, \bibinfo{person}{Yang Song},
  \bibinfo{person}{Thomas Leung}, {and} \bibinfo{person}{Ian Goodfellow}.}
  \bibinfo{year}{2016}\natexlab{}.
\newblock \showarticletitle{Improving the robustness of deep neural networks
  via stability training}. In \bibinfo{booktitle}{\emph{Proceedings of the ieee
  conference on computer vision and pattern recognition}}.
  \bibinfo{pages}{4480--4488}.
\newblock


\bibitem[\protect\citeauthoryear{Zhu, Li, Zhou, and Philip}{Zhu
  et~al\mbox{.}}{2017}]%
        {zhu2017differentially}
\bibfield{author}{\bibinfo{person}{Tianqing Zhu}, \bibinfo{person}{Gang Li},
  \bibinfo{person}{Wanlei Zhou}, {and} \bibinfo{person}{S~Yu Philip}.}
  \bibinfo{year}{2017}\natexlab{}.
\newblock \showarticletitle{Differentially private data publishing and
  analysis: A survey}.
\newblock \bibinfo{journal}{\emph{IEEE Transactions on Knowledge and Data
  Engineering}} \bibinfo{volume}{29}, \bibinfo{number}{8}
  (\bibinfo{year}{2017}), \bibinfo{pages}{1619--1638}.
\newblock


\bibitem[\protect\citeauthoryear{Zinkevich, Weimer, Li, and Smola}{Zinkevich
  et~al\mbox{.}}{2010}]%
        {zinkevich2010parallelized}
\bibfield{author}{\bibinfo{person}{Martin Zinkevich}, \bibinfo{person}{Markus
  Weimer}, \bibinfo{person}{Lihong Li}, {and} \bibinfo{person}{Alex~J Smola}.}
  \bibinfo{year}{2010}\natexlab{}.
\newblock \showarticletitle{Parallelized stochastic gradient descent}. In
  \bibinfo{booktitle}{\emph{Advances in neural information processing
  systems}}. \bibinfo{pages}{2595--2603}.
\newblock


\end{thebibliography}

\end{document}